\documentclass[twocolumn]{aastex631}

\usepackage{amsmath}
\newcommand{\dif}{\mathrm{d}}
\usepackage{bm}
\usepackage{booktabs}
\usepackage{CJKutf8}
\emergencystretch=\maxdimen       
\newcommand{\new}[1]{{#1}}

\received{January 27, 2023}
\revised{March 19, 2023}
\accepted{April 2, 2023}
\submitjournal{The Astrophysical Journal}

\shorttitle{Planetesimal IMFs under diffusion regulated collapse}
\shortauthors{Gerbig and Li}

\graphicspath{{./}{figures/}}

\begin{document}

\title{Planetesimal Initial Mass Functions following Diffusion Regulated Gravitational Collapse}

\correspondingauthor{Konstantin Gerbig}
\email{konstantin.gerbig@yale.edu}

\author[0000-0002-4836-1310]{Konstantin Gerbig}
\affiliation{Department of Astronomy, Yale University, 52 Hillhouse Ave, New Haven, CT 06511, USA}

\author[0000-0001-9222-4367]{Rixin Li \begin{CJK*}{UTF8}{gbsn}
(李日新)
\end{CJK*}}
\affiliation{Center for Astrophysics and Planetary Science, Department of Astronomy, Cornell University, Ithaca, NY 14853, USA}

\begin{abstract}
The initial mass function (IMF) of planetesimals is of key importance for understanding the initial stages of planet formation, yet theoretical predictions so far have been insufficient in explaining the variety of IMFs found in simulations. Here, we connect diffusion-tidal-shear limited planetesimal formation within the framework of a Toomre-like instability in the particle mid-plane of a protoplanetary disk to an analytic prediction for the planetesimal IMF. The shape of the IMF is set by the stability parameter $Q_\mathrm{p}$, which in turn depends on the particle Stokes number, the Toomre $Q$ value of the gas, the local dust concentration and the local diffusivity. We compare our prediction to high-resolution numerical simulations of the streaming instability and planetesimal formation via gravitational collapse. We find that our IMF prediction agrees with numerical results, and is consistent with both the `planetesimals are born big' paradigm and the power law description commonly found in simulations.
\end{abstract}


\section{Introduction} \label{sec:intro}

Planetesimals are the initial building blocks of planets and their Initial Mass Function (IMF) is consequently of great interest. The planetesimal formation process in protoplanetary disks, which ultimately dictates the planetesimal IMF, is connected to an ensemble of challenges, one of the most prominent of which is the so-called \textit{meter barrier} \new{\citep[see the reviews by][]{Chiang2010, Liu2020, Drazkowska2022}}. Particles exceeding sizes of order meters are expected to be limited in their capacity to grow via continued coagulation due to both rapid radial drift inwards and increased relative velocities that result in preferentially destructive collisions \citep[][]{Birnstiel2012}. One well-received solution to this growth barrier is to rapidly form planetesimals via gravitational collapse of over-dense particle clouds. Initial ideas focused on the gravitational instability of the entire particle mid-plane \citep[][]{Safronov1969, Goldreich1973}. However, Kelvin-Helmholtz stirring prevents razor-thin particle settling and thus renders such a global gravitational instability challenging \citep[][]{Weidenschilling1980, Sekiya1998, Youdin2002, Johansen2006KHI}. On the other hand, local patches have been shown to collapse if local particle concentrations are sufficiently high \citep[][]{Johansen2006gravturb}. In particular, instabilities energized by the relative dust-gas streaming velocity \citep[][]{Youdin2005, Squire2018} can concentrate particles to densities sufficient for gravitational collapse to trigger and planetesimals to form \citep[e.g.,][]{Johansen2015, Gerbig2020}. In the past, the IMF of planetesimals has been obtained by performing numerical simulations of this setup and then counting produced planetesimals, a process that resulted in power-laws of various kinds \citep[][]{Simon2016, Schaefer2017, Li2019}.

Recently, the formation of planetesimals has been connected to the diffusion of particles as well as their stability to stellar tidal forces \citep[][]{Klahr2016, Gerbig2020, Klahr2020, Klahr2021}. This framework envisions particles to be subject to a diffusive flux away from the concentration maximum, much how pressurized gas clouds resist collapse in star formation. In addition, any particle cloud on the verge of gravitational collapse must be stable to stellar tidal gravity. These effects limit planetesimal formation on small scales and large scales respectively, thus together prescribe a characteristic scale on which planetesimal formation is expected to occur. This in turn, can be translated to a characteristic planetesimal mass, which \citet[][]{Klahr2020} hypothesized to be the center of a Gaussian-shaped IMF, in agreement with IMFs of primordial asteroids \citep{Delbo2019}, yet seemingly implying a mismatch to numerically obtained IMFs.

In this paper, we connect these two paradigms by deriving planetesimal IMFs within the framework of diffusion-tidal-shear limited planetesimal formation. Thereby, we argue that the probability density function of a given scale to collapse scales with the scale's Toomre-like growth rate. We also directly test our prediction by conducting numerical simulations using proven setups in \texttt{ATHENA} \citep[][]{Stone2008}, that produce the streaming instability \citep{Youdin2005} and planetesimal formation. In the process, we for the first time conduct diffusion measurements in large-scale stratified streaming instability simulations, as well as develop a method for obtaining local particle concentrations that are appropriate for characterizing the onset of planetesimal formation.

The paper is structured as follows. In Sect.~\ref{sect:Toomreinstability} we review the Toomre-like instability for planetesimal formation. In Sect.~\ref{sect:IMF}, we derive planetesimal IMFs, which we compare to numerical simulations in Sect.~\ref{sect:numerical_tests}. We discuss our results, namely applicability, caveats and implications, in Sect.~\ref{sect:discussion}.

\section{Toomre-like instability}
\label{sect:Toomreinstability}

\subsection{Stability parameter}

The stability of self-gravitating particles subject to a diffusive flux induced by coupling to turbulent gas velocities has been studied in numerous occasions \citep{Youdin2011, Gerbig2020, Klahr2020, Klahr2021}, and can be assessed using a Toomre-like value $Q_\mathrm{p}$, such as
\begin{align}
\label{eq:Qp_definition}
    Q_\mathrm{p} \equiv \sqrt{\frac{\delta_r}{\mathrm{St}}}\frac{1}{Z}\frac{ c_\mathrm{s}\Omega}{ \pi G \Sigma_\mathrm{g}} = \sqrt{\frac{\delta_r}{\mathrm{St}}}\frac{Q}{Z}
\end{align}
where $Z = \Sigma_\mathrm{p}/\Sigma_\mathrm{g}$ is the (local) dust concentration, $Q = c_\mathrm{s}\Omega /(\pi G \Sigma_\mathrm{g})$ is the standard Toomre value \citep[][]{Toomre1964}, $\mathrm{St} = t_\mathrm{s}\Omega$ is the dimensionless stopping time, and $\delta_r = D_{\mathrm{p,}r}/(c_\mathrm{s}H)$ is the dimensionless (radial) diffusion coefficient for particles. $\Omega$ is the orbital frequency, $c_\mathrm{s}$ the sound-speed of the gas, $H = c_\mathrm{s}/\Omega$ the disk pressure scale-height, $\Sigma_\mathrm{p}$ and $\Sigma_\mathrm{g}$ particle and gas surface densities respectively, and $G$ is the gravitational constant. In the Epstein regime, the stopping time relates to particle size $a$, and dust material density $\rho_\bullet$ via $t_\mathrm{s} = \rho_\bullet a /(\sqrt{2\pi}\Sigma_\mathrm{g})$. We consider dust and pebbles where $t_\mathrm{s}$ is such that $\mathrm{St}$ remains below unity \citep[e.g.,][]{Birnstiel2012}. This corresponds to well or at least marginally coupled particles.

If $Q_\mathrm{p} < 1$, the system is unstable and expected to collapse and form planetesimals. We review the corresponding instability analysis in the following section.

\subsection{Dynamical equations and dispersion relation}

\begin{figure*}
    \centering
    \includegraphics[width =\linewidth]{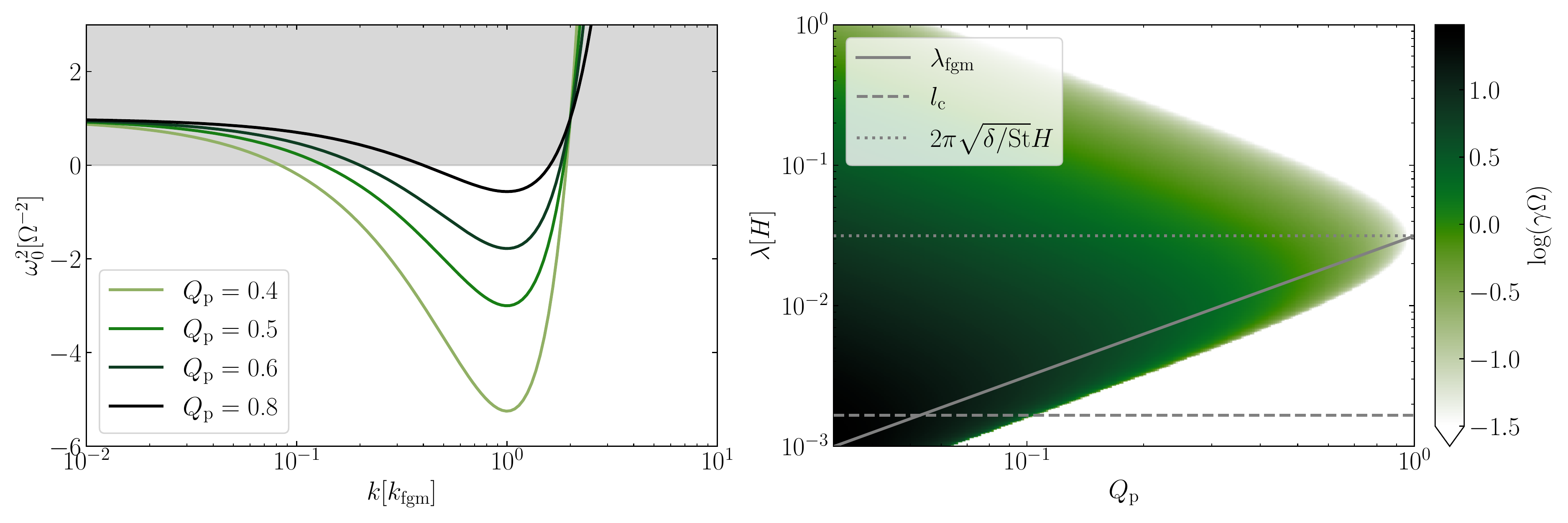}
    \caption{Dispersion relation $\omega_0^2$ in Eq.~\eqref{eq:disprel} (left panel) and growth rates $\gamma$ in Eq.~\eqref{eq:growth_rate} (right panel) for different values of $Q_\mathrm{p}$. Parameters are $\delta = 10^{-5}$ and $\mathrm{St} = 0.4$. The gray shaded region in the left panel indicates stability against axisymmetric perturbations. As $Q_\mathrm{p}$ decreases, more modes become unstable, and the fastest growing mode (dashed lines in right panel) shifts to smaller scales. For reference, we plot the scale $2\pi\sqrt{\delta/\mathrm{St}}H$ (dotted line right panel).}
    \label{fig:disprel}
\end{figure*}

Following \citep[][]{Klahr2021}, we start with the a set of dynamical equations for dust particles  in the shearing sheet approximation and under the assumption of a razor-thin disk. We adopt the coordinates $r,\phi, z$ for radial, azimuthal and vertical directions respectively, and consider a patch at distance $R$ from a solar-mass star. The linearized set of equation is given by
\begin{align}
    \frac{1}{\Sigma_\mathrm{p}}\frac{\partial \Sigma_\mathrm{p}^\prime}{\partial t} + \frac{\partial v_r^\prime}{\partial r} = 0, \\
    \label{eq:momentum_radial}
    \frac{\partial v_r^\prime}{\partial t} - 2\Omega v_\phi^\prime = - \frac{1}{\Sigma_\mathrm{p}}\frac{D_\mathrm{p,r}}{t_\mathrm{s}}\frac{\partial \Sigma_\mathrm{p}^\prime}{\partial r} - \frac{\partial \Phi^\prime}{\partial r} - \frac{v_r^\prime}{t_\mathrm{s}}, \\
    \frac{\partial v_\phi^\prime}{\partial t} + \frac{\Omega v_r^\prime}{2} = 0,
\end{align}
where the prime denotes perturbed quantities. They describe mass continuity, and conservation of radial and azimuthal momenta. Notably, we ignore an explicit azimuthal drag term as in e.g., \citet[][]{Youdin2011}, an assumption that is justified at dust-to-gas ratios of order unity. Instead, coupling to the gas is assumed to be wholly described by gas pressure counteracting radial contraction and the diffusive flux in the radial momentum equation, which can be understood as an effective pressure flux induced by turbulent gas motions. As a consequence, there is also no mass diffusion term in the continuity equation. 

We continue by introducing axisymmetric WKB waves scaling with $\Sigma_\mathrm{p}^\prime \propto \exp(-i(kr-\omega t))$. $\Phi^\prime = - 2\pi G \Sigma^\prime /|k|$ is the potential for a perturbed disk assuming $\rho_\mathrm{p}(k,z) = (k\Sigma_\mathrm{p}/2)\exp(-|k|z)$. The resulting dispersion relation is given by \citep[Eq. (B22) in][]{Klahr2021},
\begin{align}
\label{eq:disprel}
    \omega^2_0 = \frac{\delta_r}{\mathrm{St}}c_\mathrm{s}^2 k^2 - 2\pi \Sigma_\mathrm{p} G |k| +\Omega^2 ,
\end{align}
and can be expressed in terms of the stability parameter $Q_\mathrm{p}$
\begin{align}
    \frac{\omega^2_0}{\Omega^2} = \frac{\delta_r}{\mathrm{St}} (kH)^2 - \frac{2}{Q_\mathrm{p}}\sqrt{\frac{\delta_r}{\mathrm{St}}} |k|H + 1.
\end{align}
Here, $\omega_0^2$ is defined as $\omega_0^2 = \omega (\omega - i/t_\mathrm{s})$, and represents the complex frequency without the drag term in Eq.~\eqref{eq:momentum_radial}. Note that the herein used solution to the Poisson equation does not take into account a softening term caused by the particle layer thickness \citep[see Eq.~(12) in][]{Youdin2011}.

\subsection{Fastest growing mode and growth rates.}

\begin{figure*}
    \centering
    \includegraphics[width =\linewidth]{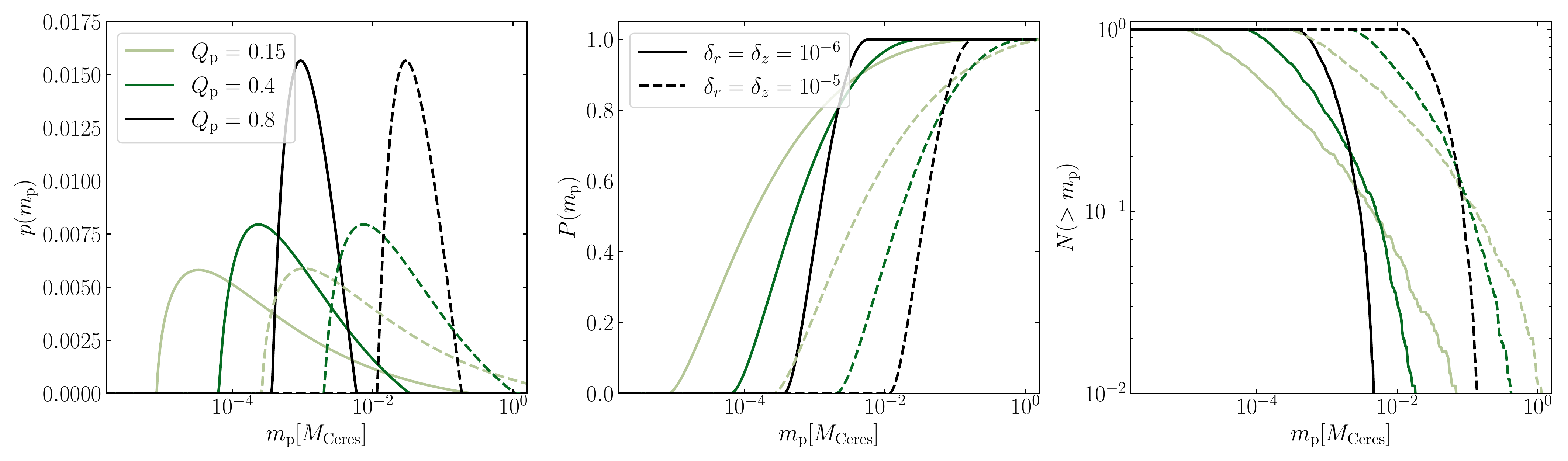}
    \caption{Probability density functions (left panel), cumulative probability density functions (center panel) and initial mass functions (right panel) for the growth rates in Fig.~\ref{fig:imf_1} where $\mathrm{St} = 0.4$. Diffusion is isotropic and set to $\delta = 10^{-6}$ (solid lines) and $\delta = 10^{-5}$ (dashed lines). Different colors correspond to different values of $Q_\mathrm{p}$, all of which chosen to be unstable. We use Eq.~\eqref{eq:massathill} for converting unstable scale to seed mass, which assumes collapse at Hill density, and requires knowledge of the aspect ratio which we set to $h = 0.05$. We assume the system to produce $N = 1000$ planetesimals total to calculate the normalized IMF $N(<m_\mathrm{p})$.}
    \label{fig:imf_1}
\end{figure*}

The fastest growing mode is found by solving $\partial \omega_0^2 / \partial k = 0$, which yields
\begin{align}
    k_\mathrm{fgm} = \sqrt{\frac{\mathrm{St}}{\delta_r}}\frac{1}{Q_\mathrm{p}H}.
\end{align}
The complex frequency of the fastest growing mode is simply
\begin{align}
    \frac{\omega^2_{0,\mathrm{fgm}}}{\Omega^2} = 1 - \frac{1}{Q_\mathrm{p}^2},
\end{align}
which highlights that only for $Q_\mathrm{p} < 1$ exist $k$ for which $\omega_0^2 < 0$ and the instability can grow, thus justifying the definition of $Q_\mathrm{p}$ as a stability parameter. We can also calculate largest and smallest unstable scales $\lambda = 2\pi/k$ by solving $\omega^2_0 = 0$, resulting in
\begin{align}
\label{eq:lambda_minmax}
    \frac{\lambda_\mathrm{min/max}}{\lambda_\mathrm{fgm}} = \frac{1}{Q_\mathrm{p}}\left(\frac{1}{Q_\mathrm{p}} \pm \sqrt{\frac{1}{Q_\mathrm{p}^2} - 1}\right),
\end{align}
with $\lambda_\mathrm{fgm} = 2\pi/k_\mathrm{fgm}$. The range of unstable scales is $\lambda_\mathrm{max} - \lambda_\mathrm{min} = 2\lambda_\mathrm{fgm}/Q_\mathrm{p}\sqrt{1/Q_\mathrm{p}^2 - 1}$, which increases for decreasing $Q_\mathrm{p}$.

Note, that the fastest growing mode  relates to the critical cloud radius $l_\mathrm{c}$ in \citet{Klahr2020} and \citet{Gerbig2020} via
\begin{align}
    \lambda_\mathrm{fgm} = 2\pi \sqrt{\frac{\delta_r}{\mathrm{St}}}Q_\mathrm{p} H  = 6 \pi Q_\mathrm{p} l_\mathrm{c}.
\end{align}

The growth rate $\gamma(k)$ can be found solving $\gamma = i\omega$ or equivalently $\gamma(\gamma +1/t_\mathrm{s}) = - \omega^2_0$ \citep[][]{Klahr2021}:
\begin{align}
\label{eq:growth_rate}
    \frac{\gamma(k)}{\Omega} = - \frac{1}{2\mathrm{St}} + \sqrt{\frac{1}{4\mathrm{St}^2} - \frac{\omega^2_0(k)}{\Omega^2}}.
\end{align}
Hence, the fastest growing mode $k_\mathrm{fgm}$ grows with
\begin{align}
    \frac{\gamma(k = k_\mathrm{fgm})}{\Omega} = - \frac{1}{2\mathrm{St}} + \sqrt{\frac{1}{4\mathrm{St}^2} + \frac{1}{Q_\mathrm{p}^2}- 1}
\end{align}
Dispersion relation and growth rates are shown for a set of unstable $Q_\mathrm{p}$ in Fig.~\ref{fig:disprel}. In the right panel, we also plot $\lambda_\mathrm{fgm}, l_\mathrm{c}$ as well as the scale $2\pi \sqrt{\delta_r/\mathrm{St}}H$ which is of order the radial extent of small-scale particle structures \citep{Gerbig2020}.


\section{Planetesimal initial mass functions}
\label{sect:IMF}

\subsection{Planetesimal masses at Hill density}
\label{sect:plsmass_athill}

In order for a region in the particle disk to be stable against stellar tidal gravity, its mass must be be contained within its Hill-radius. In other words, the region's density must be at least Hill density 
\begin{align}
\label{eq:hilldens}
    \rho_\mathrm{H} = \frac{9 \Omega^2}{4\pi G}.
\end{align}

Assuming a given scale $k = 2\pi/\lambda$ is unstable under the previously discussed dispersion relation, at Hill density $\rho_\mathrm{H}$, the mass available to the produced planetesimal\footnote{The mass $m_\mathrm{p}$ is only equal to the initial planetesimal mass if the entire particle cloud collapses to material density of the planetesimal, and thus should be understood as an approximate mass scale. As such, in \citet[][]{Klahr2020, Klahr2021}, this mass scale is called `equivalent mass', and associated with some conversion efficiency that quantifies the fraction that ultimately ends up in the formed planetesimal.} can be estimated with 
\begin{align}
    m_\mathrm{p} = \frac{\pi}{4}\left(\frac{\lambda}{2\pi}\right)^2 \Sigma_\mathrm{p}(\rho_\mathrm{p} = \rho_\mathrm{H})
\end{align}
We assumed that the seed mass has access to a region of size $\lambda/2\pi$. The surface density if the mid-plane has Hill-density can be estimated with $\Sigma_\mathrm{p} = \sqrt{2\pi}H_\mathrm{p}\rho_\mathrm{p}$, where the particle scale height relates to the vertical diffusion coefficient via \citep{Youdin2007}
\begin{align}
\label{eq:particle_scaleheight}
    H_\mathrm{p} = \sqrt{\frac{\delta_z}{\mathrm{St}}} H.
\end{align}
\new{Due to anisotropic diffusion where $\delta_r \neq \delta_z$, contraction does not necessarily proceed spherically symmetric \citep[compare to][]{Klahr2021}.}

\new{Assuming a solar mass star}, the mass associated with an unstable scale $\lambda$ is of order,
\begin{align}
\label{eq:massathill}
   m_\mathrm{p} =\frac{9}{64}\sqrt{\frac{2}{\pi^3}} h^3 \sqrt{\frac{\delta_z}{\mathrm{St}}} \left(\frac{\lambda}{H}\right)^2 M_\odot,
\end{align}
and equivalently, the unstable scale $\lambda$ that is expected to produce a planetesimal of mass of order $m_\mathrm{p}$ is
\begin{align}
\label{eq:lambdaofmass}
    \frac{\lambda}{H} = \left(\frac{64}{9}\sqrt{\frac{\pi^3}{2}}\frac{1}{h^3}\sqrt{\frac{\mathrm{St}}{\delta_z}}\frac{m_\mathrm{p}}{M_\odot}\right)^{1/2}
\end{align}
Here $h = H/R$ is the disk aspect ratio, typically of order $0.03 < h < 0.1$. The mass associated with the fastest growing mode is given by
\begin{align}
    m_\mathrm{p,fgm} = \frac{9}{8}\sqrt{\frac{\pi}{2}}\sqrt{\frac{\delta_z}{\mathrm{St}}}\frac{\delta_r}{\mathrm{St}} h^3Q_\mathrm{p}^2  M_\odot
\end{align}
which scales as
\begin{align}
\label{eq:mpfgm_scaling}
\begin{split}
    m_\mathrm{p,fgm} &\approx 0.37 M_\mathrm{Ceres} \cdot \left(\frac{\delta_z}{10^{-5}}\right)^{\frac{1}{2}} \left(\frac{\delta_r}{10^{-5}}\right) \\ & \cdot \left(\frac{0.1}{\mathrm{St}}\right)^{\frac{3}{2}}\left(\frac{h}{0.05}\right)^3 \left(\frac{Q_\mathrm{p}}{1}\right)^2
\end{split}
\end{align}
Under isotropic diffusion, this equals the characteristic mass in \citet{Klahr2020} if $Q_\mathrm{p} \approx 0.28$. \new{We can also compare this mass scaling to \citet{Liu2020_planetform} who comprise numerical results \citet{Johansen2015, Simon2016, Schaefer2017, Abod2019, Li2019} into a characteristic mass which, for their fiducial parameters and a solar mass star likewise yields $\sim 0.3 M_\mathrm{Ceres}$ and is thus consistent with Eq.~\eqref{eq:mpfgm_scaling}.}

Note, that the Toomre instability a priori does not require Hill-density to operate. Indeed, as pointed out by \citet{Klahr2021}, if the vertical density structure is not set by vertical diffusion, but instead stringently follows $\rho_\mathrm{p}(k,z) = (k\Sigma_\mathrm{p}/2)\exp(-|k|z)$, then the fastest growing linear instability would be achieved at a mid-plane density of $\rho_\mathrm{p}(k = k_\mathrm{fgm}, z = 0) = 2\rho_\mathrm{H}/9$. By presupposing Hill density, our approach detaches from this assumption of mode-dependent stratification, and in the process excludes Toomre unstable clouds that fail to withstand tidal gravity and are thus of little physical importance for the IMF.

\subsection{IMF from Toomre growth rate}
\label{sect:imf1}

Given this context of Toomre-like instability and Hill density, we proceed by providing predictions for the IMF.
Our ansatz is to take the mode-depended instability growth rates as a probability density function $p(\lambda)$ for unstable scales, and then convert unstable modes to planetesimal masses via the recipes discussed in Sect.~\ref{sect:plsmass_athill}. This yields a probability density function $p(m_\mathrm{p})$ such that the probability of a seed mass within $[m_\mathrm{p}, m_\mathrm{p}+\dif m_\mathrm{p}]$ is $p(m_\mathrm{p})\dif m_\mathrm{p}$. This first order approach agrees with intuition in that fastest growing modes should most preferentially collapse, slowly growing modes only sometimes, stable modes never.  

We write the probability density function $p(\lambda)$ for scale $\lambda$ to collapse as
\begin{align}
    p(\lambda)  \propto \mathrm{max}[\hat{\gamma}^{-1} \gamma(\lambda), 0]
\end{align}
$\hat{\gamma}^{-1}$ is a normalization constant given by $ \hat{\gamma} = \int_{\lambda_\mathrm{min}}^{\lambda_\mathrm{max}}\gamma(\lambda) \dif \lambda$. Since $\lambda \propto m_\mathrm{p}^{1/2}$ via Eq.~\eqref{eq:lambdaofmass}, we can map $p(\lambda)\dif\lambda$ onto the probability density function for seed planetesimal masses $p(m_\mathrm{p})\dif m_\mathrm{p}$.

The probability density function $p(m_\mathrm{p})$, the cumulative probability function $P(m_\mathrm{p}^\prime \leq m_\mathrm{p})$, and the resulting IMF are shown in Fig.~\ref{fig:imf_1}, for $\mathrm{St} = 0.4$ and different values of $\delta$ and $Q_\mathrm{p}$. The smaller $Q_\mathrm{p}$, the flatter the IMF, as more scales become unstable. The steepest IMF is achieved for $Q_\mathrm{p} \rightarrow 1$. Indeed, for $Q_\mathrm{p} = 1$, the probability density function becomes a Dirac delta function, i.e., $p(m_\mathrm{p}) = \delta(m_\mathrm{p,H}- m_\mathrm{p})$. Increasing diffusivity shifts the IMF to larger masses, provided $Q_\mathrm{p}$ remains constant, which would require a corresponding decrease in $Q/Z$.

\subsection{Comparison to statistical approaches}

Our ansatz is distinct from past means of deriving planetesimal IMFs. \citet{Cuzzi2008, Cuzzi2010, Hartlep2020} approach the problem statistically, and consider turbulent clustering of particles. In particular, the assumed scale invariance of the turbulent spectrum implies the statistically appearance of regions of highly enhanced particle density. The argument is that only sufficiently dense clumps can withstand ram pressure disruption and thus contract to planetesimals in a process called \textit{primary accretion}, thus limiting the formation of planetesimals on the low-mass end. Our mechanism likewise prohibits the formation of arbitrarily small planetesimals, yet the physical intuition differs in that (1) the system is, in fact, gravitationally unstable under the Toomre-like instability discussed in Sect.~\ref{sect:Toomreinstability} resulting in (2) ram pressure being negligible compared to diffusion (also see \citet{Klahr2020}), and (3), that the smallest planetesimals are those resulting from the smallest scale that can be gravitationally unstable under the dispersion relation in Eq.~\eqref{eq:disprel}. 

Another statistical ansatz was taken by \citet{Hopkins2013}, where turbulent density fluctuations can render local regions of the (gas) disk unstable to gravitational collapse. The initial mass function of collapsing clumps therein depends on the critical density for self-gravitating clumps and as well as the properties of the ambient turbulence. In our ansatz, the gas remains stable throughout. Moreover, we take all clumps to collapse exactly at Hill density in Eq.~\eqref{eq:hilldens}. Different masses are the result of collapse of differently sized regions, i.e. on different scales $\lambda$.

\section{Numerical Tests}
\label{sect:numerical_tests}

We perform high-resolution numerical tests using \texttt{ATHENA} \citep[][]{Stone2008, Bai2010} to directly test our predictions for the IMF. The procedure is similar to that in e.g.~\citet[][]{Li2019, Gerbig2020}, in that we let a small patch around a protoplanetary disk mid-plane evolve into some turbulent state. We then turn on self-gravity with different values for $Q_\mathrm{p}$ by varying the self-gravity parameter $\hat{G}$ in code units. The measurement of $Q_\mathrm{p}$, as well as the calculation of the predicted IMF requires measurement of radial and vertical diffusion, prior to turning on-self gravity. Since the streaming instability will concentrate particles into filaments, we must also determine the local particle concentration.

\subsection{Numerical Setup}

\begin{figure}
    \centering
    \includegraphics[width = \linewidth]{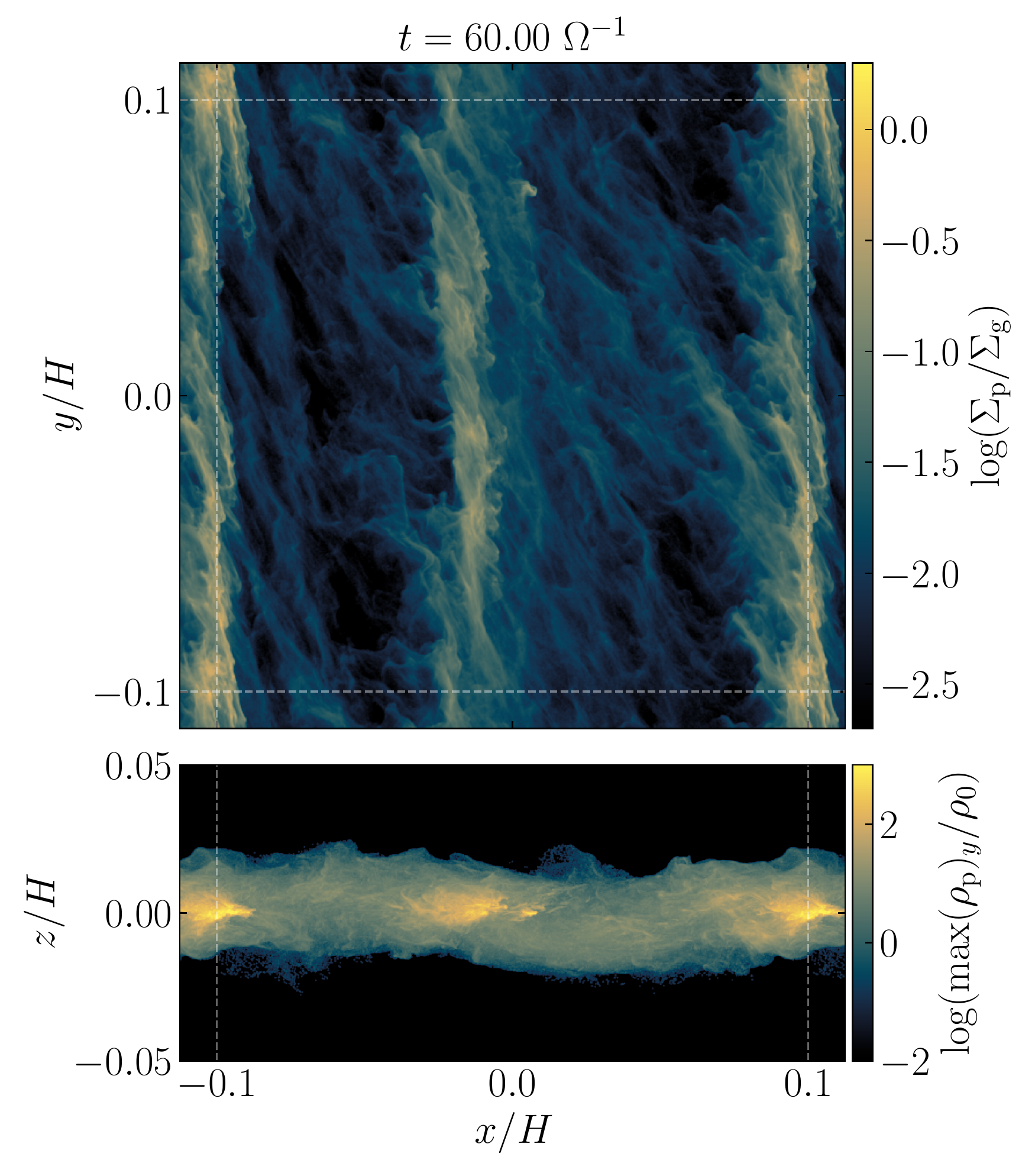}
    \caption{Map of the particle surface density at after 60 $\Omega^{-1}$, which is when self-gravity is turned on. Top panel is vertically integrated, wheares bottom panel is azimuthally integrated. The white dashed lines mark the border of the physical simulation domain and the ghost cells. Since, one of the two filaments at 60 $\Omega^{-1}$ is located right at the radial simulation boundary, we chose to include the ghost cells in this figure, which results in that filament being depicted twice.}
    \label{fig:snapshot}
\end{figure}

\begin{figure*}
    \centering
    \includegraphics[width = \linewidth]{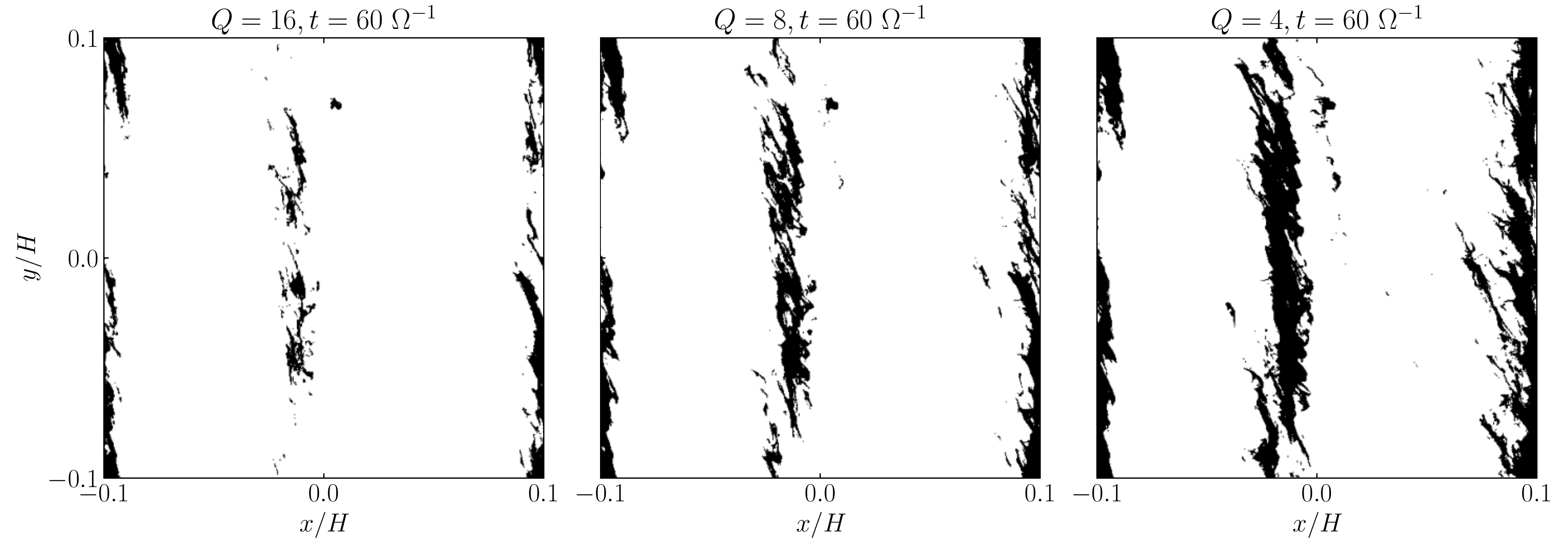}
    \includegraphics[width = \linewidth]{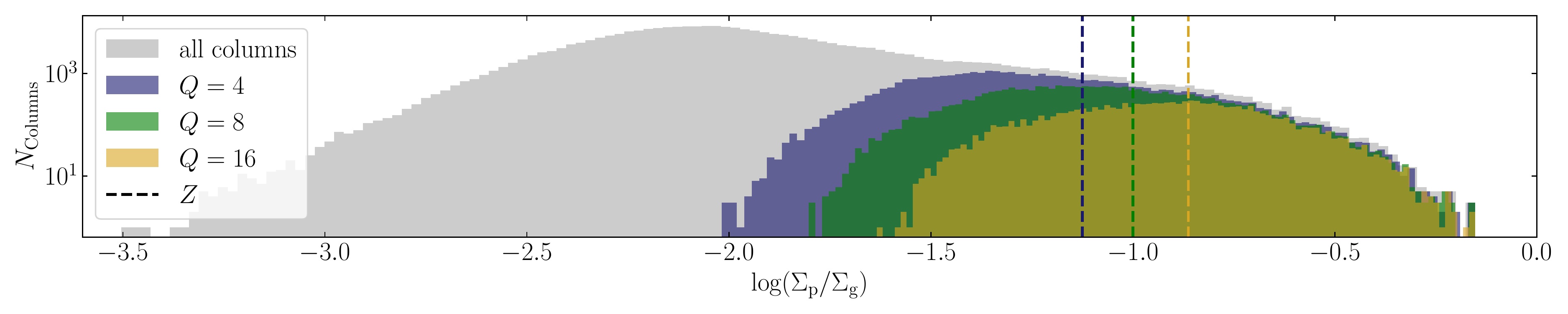}
    \caption{Maps of the columns that, at 60 $\Omega^{-1}$, contain at least one cell at or above Hill-density for $Q = 16$, $Q = 8$, and $Q = 4$ (top panels, from left to right), and corresponding column density histograms (bottom panel). The bottom panel also indicates the chosen values for $Z$ in dashed lines, i.e. $Z = 0.137 $, $Z = 0.100 $, and $Z = 0.0075$ for $Q = 16$, $Q = 8$, and $Q = 4$ respectively. The full surface density map of this snapshot is shown in Fig.~\ref{fig:snapshot}.  }
    \label{fig:concentration_measure}
\end{figure*}

We employ \texttt{ATHENA} \citep[][]{Stone2008} to solve the hydrodynamic equations on an Eulerian grid including Lagrangian super-particles \citep[][]{Bai2010}. Our numerical setup is similar to simulations of streaming instability regulated planetesimal formation such as \citet{Johansen2007Nature,Simon2016,Li2019,Gerbig2020}, in that we use the local shearing box approximation \citep{Goldreich1965} with coordinates $(x,y,z)$. We consider a non-magnetized gas with an isothermal equation of state. Gas is initialized in hydrostatic equilibrium.  Particles are forced into a super-Keplerian rotation by an external pressure gradient, which we parameterize using
\begin{align}
    \Pi = \frac{\eta v_\mathrm{K}}{c_\mathrm{s}},
\end{align}
where $\eta = - (1/2)h^2 \dif \ln \rho / (\dif \ln r)$. This is mathematically equivalent  to real disks where particles orbit Keplerian, and gas experiences sub-Keplerian forcing  (see \citealt{Bai2010} for details on the pressure gradient implementation in \texttt{ATHENA}). Note, that $\Pi$ relates to the pressure gradient parameter used in \citet[][]{Schreiber2018, Gerbig2020} simply via $\Pi = \beta/2$. Particles are initialized with a narrower Gaussian, however at an scale height of $\eta r/ 2$, which is the characteristic scale of Kelvin-Helmholtz instability in protoplanetary disks \citep[][]{Gerbig2020}. As such, particles are not expected to undergo significant settling or lofting due to vertical stellar gravity or Kelvin-Helmholtz stirring respectively. 

The pressure gradient and the resulting relative velocity between particle and gas flow energize the linear streaming instability \citep[][]{Youdin2005, Squire2018}, which then saturates non-linearly and in the process concentrates particles into high density regions \citep{Johansen2007}. The degree of which the streaming instability operates, and as a consequence the strength of particle diffusion and concentration prior to gravitational collapse and planetesimal formation, is largely set by two quantities. First, the Stokes number $\mathrm{St}$, and second, the ratio of metallicity and pressure gradient $Z_0/\Pi$ \citep[][]{Sekiya2018}. The former quantifies the coupling of particles to the gas flow, in particular the gas turbulence. The latter traces the ratio of dust abundance relative to gas and dust layer scale-height, and thus maps onto the mid-plane dust-to-gas ratio. Note, that the metallicity $Z_0$ is the global (as in simulation domain averaged) particle concentration and thus not necessarily equal to the local enhancement $Z$ we introduced in Eq.~\eqref{eq:Qp_definition}. We elaborate on this important difference in Sect.~\ref{sect:concentration_measure}.

In this work, we choose $\mathrm{St} = 0.4$, $Z_0 = 0.02$ and $\Pi = 0.05$. This setup is specifically designed to provide favorable conditions for the streaming instability and, provided $Q_\mathrm{p} < 1$, planetesimal formation, in order to produce a large number of planetesimals which allows for a more statistically robust determination of the numerical IMF.  Our simulations use a computational domain of $0.2H\times0.2H\times0.15H$, with a resolution of $2560/H$ (i.e., $\Delta x \approx 3.9\cdot 10^{-4}H$) and $N_{\rm par} = 2^{26}\approx 6.71\cdot 10^{7}$ particles.  The vertical extent is slightly reduced from a cube to mitigate computational costs and is still tall enough because the particle layer remains thin all the time.  Moreover, in the vertical direction, we adopt outflow boundary conditions that are known to reduce boundary artifacts, especially in shorter boxes \citep{Li2018}.  In the radial and azimuthal directions, the standard shearing-periodic boundary conditions are imposed.

Following the precedent set by many past works of streaming instability, we express the results of our scale-free simulations in the dimensionless unit system of dynamical timescale $\Omega^{-1}$, $H$, and $\rho_0$. The simulation is run for $t = 60 \Omega^{-1}$ without self-gravity. This allows for sufficient amount of time in the non-linear phase of the streaming instability to measure diffusion.

Figure.~\ref{fig:snapshot} shows the vertically (top panel) and azimuthally (bottom panel) integrated particle densities at $60 \Omega^{-1}$, i.e. just before self-gravity is turned on. The streaming instability in a stratified disk collects particles into two azimuthally elongated filaments, which are enhanced in particle density relative to the prescribed average of $Z_0 = 0.02$.

Following the snapshot at 60 $\Omega^{-1}$ depicted in Fig.~\ref{fig:snapshot} we turn on self-gravity \new{for the particles}, the strength of which is parameterized by the self-gravity parameter $\tilde{G} \equiv 4\pi G\rho_0/\Omega^2$, which relates to $Q$ via $Q = 4/(\sqrt{2\pi} \tilde{G})$. Self-gravity is required for the concept of Hill-density to be meaningful. Specifically, the Hill-density depends on the self-gravity parameter via
\begin{align}
    \frac{\rho_\mathrm{H}}{\rho_0} = \frac{9}{\tilde{G}} = 9\sqrt{\frac{\pi}{8}} Q
\end{align}
We conduct three self-gravity runs with $Q= 16, Q = 8$ and $Q = 4$ which correspond to Hill densities of $\rho_\mathrm{H}/\rho_0 \sim 90, \rho_\mathrm{H}/\rho_0 \sim 45$ and $\rho_\mathrm{H}/\rho_0 \sim 23$ respectively.
Note that we can also now associate a Hill-radius $r_\mathrm{H}$ with a given mass $m$, i.e.,
\begin{align}
\label{eq:hill_radius}
    r_\mathrm{H} = R\left(\frac{m}{M_\odot}\right)^{\frac{1}{3}} =  \left(\frac{ \Tilde{G} m}{4\pi \rho_0}\right)^{\frac{1}{3}} = \left(\frac{9}{4\pi}\frac{m}{\rho_\mathrm{H}}\right)^{\frac{1}{3}}.
\end{align}

\subsection{Local particle concentration}
\label{sect:concentration_measure}

The stability parameter $Q_\mathrm{p}$ depends on the local particle concentration $Z = \Sigma_\mathrm{p}/\Sigma_\mathrm{g}$. This is importantly not equal to the initial, global particle concentration $Z_0$, which in our case is set to $Z_0 = 0.02$.  The reason for this lies within 
predominanetly radial concentration of particle surface density within two filaments as evident in Fig.~\ref{fig:snapshot}, where the typical particle column density exceeds $Z=0.02$ by between one and two orders of magnitude. This challenge in applying the diffusion-limited collapse criterion to proven shearing box simulation simulations of the streaming instability was also recognized in \citet[][]{Gerbig2020}, where the radial extent of streaming instability filaments was related to a radial enhancement. While this argument was sufficient to demonstrate the general applicability of a diffusion-limited collapse criterion to simulations, it lacks the precision to reliably predict the $Q_\mathrm{p}$ as it by construction cannot account for additional azimuthal enhancement within filaments.

Because of this, in this work, we choose a different pathway and utilize the requirement of Hill-density for planetesimal formation to measure the average particle concentration of the cells that are expected to participate in planetesimal formation. More precisely, in order for a given vertical column to be counted towards the metallicity, it must contain at least one cell at or above Hill density.

Figure.~\ref{fig:concentration_measure} shows maps of the columns that satisfy this requirement for all three $Q$-values. Note, we are here depicting the same snapshot at $60 \Omega^{-1}$ as in Fig.~\ref{fig:snapshot}, i.e. before self-gravity has impacted the simulation. We are merely evaluating which columns satisfy our requirement for a given Hill-density. A comparison to Fig.~\ref{fig:snapshot} confirms that this scheme constrains the metallicity measurement to the over-dense filaments only. In the bottom panel of Fig.~\ref{fig:concentration_measure}, we show histograms of the surface density of both the entire simulation domain in grey, as well only those columns that contain at least one cell above Hill density. The dashed lines indicate the average particle concentration for the three runs given, and correspond to $Z = 0.137 $, $Z = 0.100 $, and $Z = 0.075$ for $Q = 16$, $Q = 8$, and $Q = 4$ respectively.

\subsection{Radial diffusion}
\label{sect:diff_measure}

\begin{figure}
    \centering
    \includegraphics[width = \linewidth]{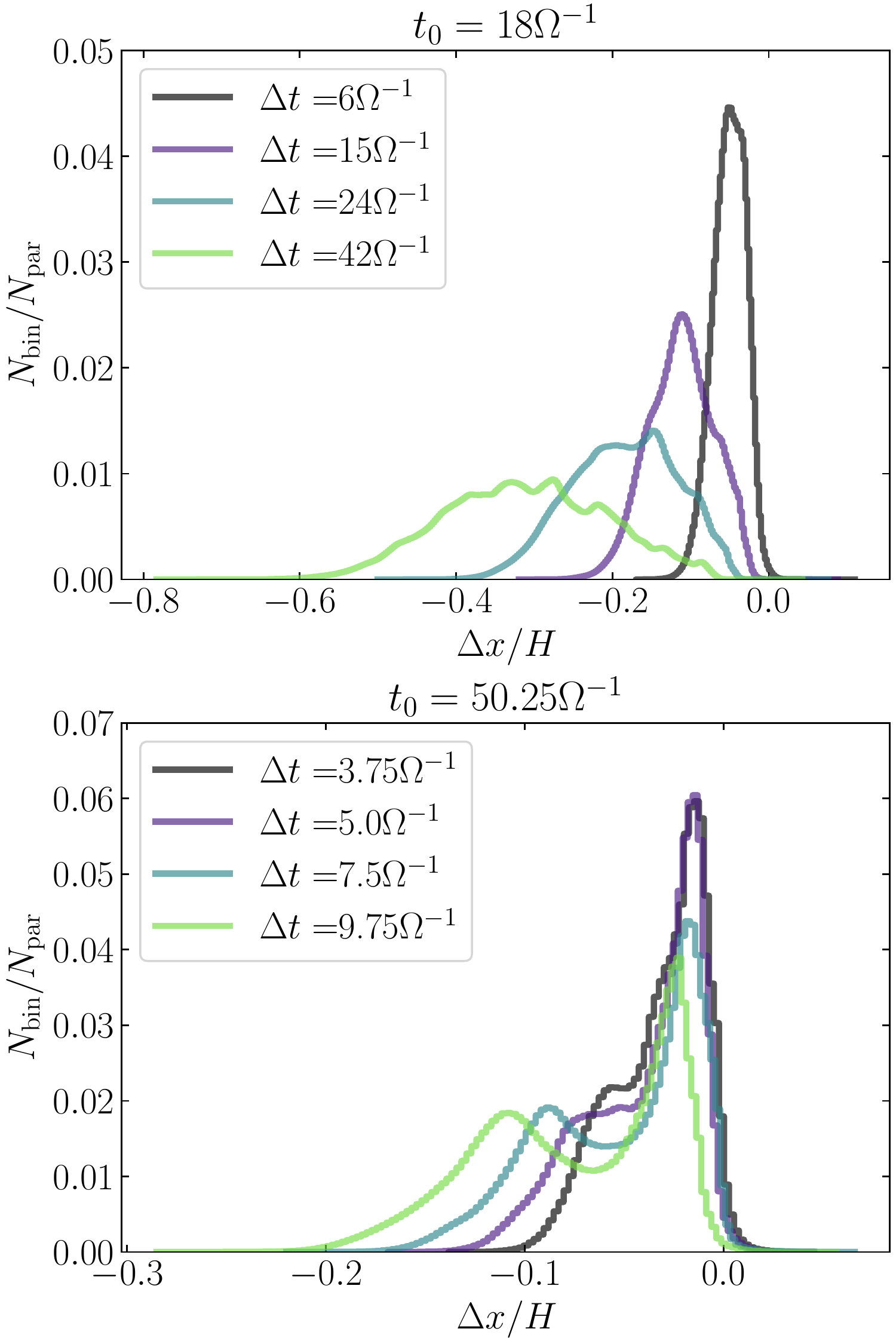}
    \caption{Evolution of an initial particle distribution at different times $\Delta t$ for $t_0 = 18 \Omega^{-1}$ (top panel) and $t_0 = 50.25 \Omega^{-1}$ (bottom panel). Due to the two over-dense filaments the particle distribution develops bi-modality --- an effect not seen in unstratified streaming instability simulations \citep[e.g.,][]{Johansen2007}
    }
    \label{fig:diffusion_measure}
\end{figure}

\begin{figure*}
    \centering
    \includegraphics[width = \linewidth]{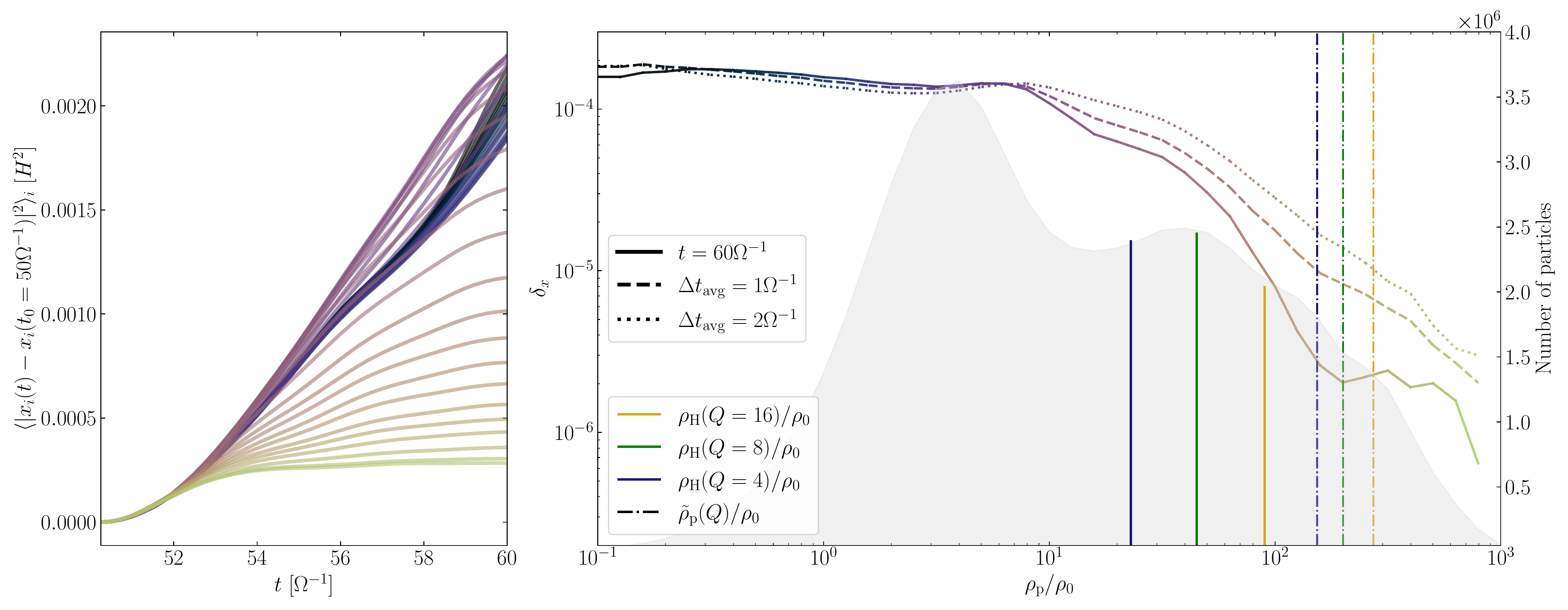}
    \caption{Radial diffusion depends on particle density. Left panel: Evolution of the variance of the particle distribution for different densities. Colors correspond to densities. Right panel: diffusion vs particle density. Colors correspond to density and map onto the left panel. Different line styles indicate different averaging times when calculating the gradient in Eq.~\eqref{eq:diffmeasure}. We overlay a histogram of the density distribution of the particles, where vertical solid lines indicate Hill density for chosen $Q$-values. Vertical dashed-dotted lines correspond to the weighted average density $\Tilde{\rho}_\mathrm{p}(Q)$, which pinpoint the diffusivities to $\delta_r = 2.4\cdot 10^{-6}$, $\delta_r = 2.0\cdot 10^{-6}$, and $\delta_r = 2.6\cdot 10^{-6}$ for $Q = 16$, $Q = 8$, and $Q = 4$ respectively}
    \label{fig:diffusion_measurebig}
\end{figure*}

We measure radial particle diffusion via \citep[][]{Youdin2007, Johansen2007, Schreiber2018, Baehr2022}
\begin{align}
\label{eq:diffmeasure}
    \frac{\delta_r}{c_\mathrm{s}H} = D_{\mathrm{p,}r} = \frac{1}{2}\frac{\partial \sigma_r^2}{\partial t},
\end{align}
The idea is, that as the turbulent state evolves particles, the underlying diffusion will widen \new{the variance $\sigma_r^2 = \langle |x_i(t) - x_i(t_0)|^2 \rangle_i$ of the radial particle distribution}, and as such allow for a calculation of a value for particle diffusion. Note, that since we are investigating the streaming instability in stratified disks, this method cannot be used to determine vertical diffusivity which we will measure in the subsequent section. 

Figure.~\ref{fig:diffusion_measure} shows the evolution of the initial particle distribution from  $18 \Omega^{-1}$ to $60 \Omega^{-1}$, when we turn on self-gravity. Our particle distribution depicts much stronger non-Gaussianity than \citet{Johansen2007} Fig.~17, since our simulation is stratified. This leads the streaming instability to produce as azimuthally-extended filaments, where particles drift slower due to the enhanced local dust-to-gas ratio. This can lead to multiple peaks in the  particle distribution --- in our specific case there are two peaks as there are two dominant filaments.

If we take all particles into account when evaluating Eq.~\eqref{eq:diffmeasure}, we will determine a global value for the radial diffusivity. However, the strength of diffusion is expected to vary locally and correlate with particle density \citep[see e.g.,][]{Schreiber2018}. As such, we sort all particles by local density and measure the evolution of the distribution up to $60 \Omega^{-1}$ for each density bin, and then calculate the corresponding diffusion coefficients via Eq.~\eqref{eq:diffmeasure}. Fig.~\ref{fig:diffusion_measurebig} shows that denser regions are less diffusive than less dense regions. More specifically, up to a dust-to-gas ratio of order $10^1$, the diffusion is approximately constant at $\delta_r = 1.3 \cdot 10^{-4}$. This is a value consistent with that found in \citet[][]{Li2021}. 

For more dense regions, the diffusion decreases significantly down to $\delta \sim 10^{-6}$ for $\rho_\mathrm{p}/\rho_0 > 4\cdot 10^2$. This turbophoretic behavior of the particles \citep[][]{Caporaloni1975, Belan2014} makes it challenging to pinpoint an exact value of $\delta$ that is appropriate for the determination of $Q_\mathrm{p}$ and the subsequent IMF. Additionally, calculation of $\delta$ requires temporal averaging. In Fig.~\ref{fig:diffusion_measurebig} we show multiple averaging times in order to highlight the trend of longer averaging times yielding larger diffusivity. 

In this work, we choose diffusivities by evaluating $\delta(\rho_\mathrm{p})$ at the particle density weighted average
\begin{align}
    \tilde{\rho}_\mathrm{p}(Q) = \frac{\sum_{i, \rho_{\mathrm{p},i} > \rho_\mathrm{H}(Q)} N_i \rho_{\mathrm{p}, i}}{\sum_{i, \rho_{\mathrm{p},i} > \rho_\mathrm{H}(Q)}}.
\end{align}
Note that we only count those particles in Hill-stable regions. The resulting densities are indicated in the right panel of Fig.~\ref{fig:diffusion_measurebig}. We evaluate the radial diffusion coefficient $\delta_r$ (see Eq. \ref{eq:diffmeasure}) with a second-order one-side derivative at $t = 60/\Omega$ (by supplying \texttt{edge\_order=2} to \texttt{numpy.gradient}). 
This results in diffusivities of $\delta_r = 2.4\cdot 10^{-6}$, $\delta_r = 2.0\cdot 10^{-6}$, and $\delta_r = 2.6\cdot 10^{-6}$ for $Q = 16$, $Q = 8$, and $Q = 4$ respectively. 

We acknowledge that due to the steepness of $\delta_x(\rho_\mathrm{H}/\rho_0)$ at high dust-to-gas ratios, as well as the dependence on averaging time $\Delta t_\mathrm{ave}$, there remains a relatively large ambiguity about the most appropriate values for $\delta_r$. Indeed, we find any diffusivity within $1 \cdot 10^{-6} \lesssim \delta_x \lesssim 1 \cdot 10^{-5}$ justifiable, and note that this full range is  consistent with diffusivities obtained by \citet{Schreiber2018} at similar dust-to-gas ratios (although with smaller particles). 

\subsection{Vertical diffusion}
\label{sect:vert_diff_measure}

\begin{figure*}
    \centering
    \includegraphics[width = \linewidth]{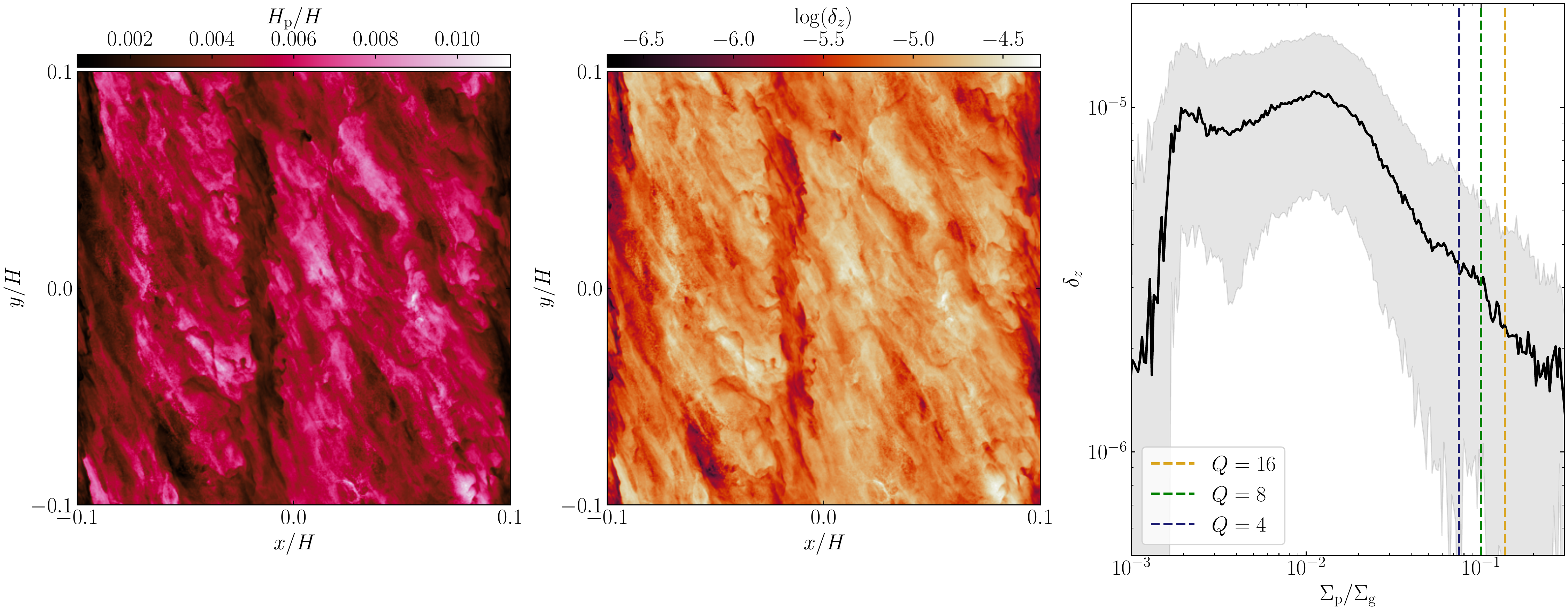}
    \caption{Map of particle scale height $H_\mathrm{p}$(left panel) and vertical diffusion coefficient $\delta_z$(center panel) at 60 $\Omega^{-1}$. The right panel shows the mean vertical diffusion $\overline{\delta}_z$ vs the local particle concentration (compare to top panel of Fig.~\ref{fig:snapshot}). The shaded region characterizes $\overline{\delta}_z \pm \sigma_\delta$, where $\sigma_\delta$ is the standard deviation of the vertical diffusivities at a given $\Sigma_\mathrm{p}/\Sigma$. The vertical lines indicate the the determined values for $Z$ (see Sect.~\ref{sect:concentration_measure}, Fig.~\ref{fig:concentration_measure}), which set the chosen values for vertical diffusivities to $\delta_z = 2.3\cdot 10^{-6}$,  $\delta_z = 3.1\cdot 10^{-6}$, and  $\delta_z = 3.3\cdot 10^{-6}$ for  $Q = 16$, $Q = 8$, and $Q = 4$ respectively.}
    \label{fig:vert_diffusion}
\end{figure*}

While vertical diffusion is not required to calculate the stability parameter $Q_\mathrm{p}$ which arises from a purely 2D consideration, it is required to calculate planetesimal masses by setting the surface density necessary to achieve tidal-stability. We first measure the particle scale-height in each vertical column of grid cells by calculating the standard deviation of the vertical particle positions, and then convert to vertical diffusivity $\delta_z$ using Eq.~\eqref{eq:particle_scaleheight}. Similar procedures were also done by e.g., \citet[][]{Bai2010plsform, Yang2018, Li2021} when analyzing the vertical diffusivity within a stratified particle layer subject to the streaming instability. The resulting maps are seen in the left and center panel of Fig.~\ref{fig:vert_diffusion}. The right panel of Fig.~\ref{fig:vert_diffusion} depicts the mean vertical diffusivity plotted vs local particle concentration, where the gray shaded region indicates one standard deviation from the mean. While there is significant spread of diffusivity for a given particle concentration, the vertical diffusivity behaves similar to the radial diffusivity in that it decreases in more concentrated regions. While in the plateau region, the diffusion anisotropy is of order $\delta_r/\delta_z \sim 10$ which is consistent with e.g., Fig.~6 in \citet[][]{Li2021}, in denser regions the diffusion is of order isotropic. Indeed, we evaluate $\delta_z(\Sigma_\mathrm{p}/\Sigma)$ at the particle concentrations identified in Sect.~\ref{sect:concentration_measure}, which yields
 $\delta_z = 2.3\cdot 10^{-6}$,  $\delta_z = 3.1\cdot 10^{-6}$, and  $\delta_z = 3.3\cdot 10^{-6}$ for  $Q = 16$, $Q = 8$, and $Q = 4$ respectively.

This relatively low, and isotropic diffusion is consistent with high particle concentrations damping diffusion from gas, regardless of direction \citep{Ida2021}.

\subsection{Planetesimals and mass scalings}
\label{sect:num_imfs}

\begin{table}[t]
    \centering
    \caption{Simulation parameters for the three self-gravity runs. For all simulations, $\mathrm{St} = 0.4, Z_0 = 0.02$, $\Pi = 0.05$, and self-gravity is turned on after $60 \Omega^{-1}$. Calculated masses assume $h = 0.05$ and collapse at Hill-density.  Our simulations have a radial and azimuthal domain size of $0.2 H$, and the smallest resolved scale is $\Delta x = 3.9\times 10^{-4}H$. The first two rows are prescribed values. Rows three, four and five correspond to quantities measured at $60 \Omega^{-1}$ as described in Sects.~\ref{sect:concentration_measure}, \ref{sect:diff_measure} and \ref{sect:vert_diff_measure}. Subsequent rows are properties calculated within the diffusion-limited collapse framework outlined in Sects.~\ref{sect:Toomreinstability} and \ref{sect:IMF}. }
\begin{tabular}{lcccl}\toprule
$Q$ & 16 & 8 & 4 \\
$\rho_\mathrm{H}/\rho_0$ & 90.2 & 45.1 & 22.6 \\\midrule
$Z$ & 0.137 & 0.100 & 0.075 \\
$\delta_r$ & $2.42 \cdot 10^{-6}$ & $2.04 \cdot 10^{-6}$ & $2.60 \cdot 10^{-6}$ \\
$\delta_z$ & $2.35 \cdot 10^{-6}$ & $3.07 \cdot 10^{-6}$ & $3.28 \cdot 10^{-6}$\\ \midrule
$Q_\mathrm{p}$ & 0.287 & 0.180 & 0.136 \\
$\lambda_\mathrm{min}[H]$ & $2.26 \cdot 10^{-3}$ & $1.29 \cdot 10^{-3}$ & $1.09 \cdot 10^{-3}$ \\
$\lambda_\mathrm{fgm}[H]$ & $4.42 \cdot 10^{-3}$ & $2.56 \cdot 10^{-3}$ & $2.18 \cdot 10^{-3}$ \\
$\lambda_\mathrm{max}[H]$ & $1.05 \cdot 10^{-1}$ & $1.56 \cdot 10^{-1}$ & $2.34 \cdot 10^{-1}$ \\
$m_\mathrm{p,min}[M_\mathrm{Ceres}]$ & $1.16 \cdot 10^{-4}$ & $4.32 \cdot 10^{-5}$ & $3.22 \cdot 10^{-5}$ \\
$m_\mathrm{p,fgm}[M_\mathrm{Ceres}]$ & $4.44 \cdot 10^{-4}$ & $1.70 \cdot 10^{-4}$ & $1.27 \cdot 10^{-4}$ \\
$m_\mathrm{p,max}[M_\mathrm{Ceres}]$ & 0.253 & 0.630 & 1.47\\\bottomrule 
\end{tabular}
    \label{tab:sim_values}
\end{table}

\begin{figure*}[t]
    \centering
    \includegraphics[width = \linewidth]{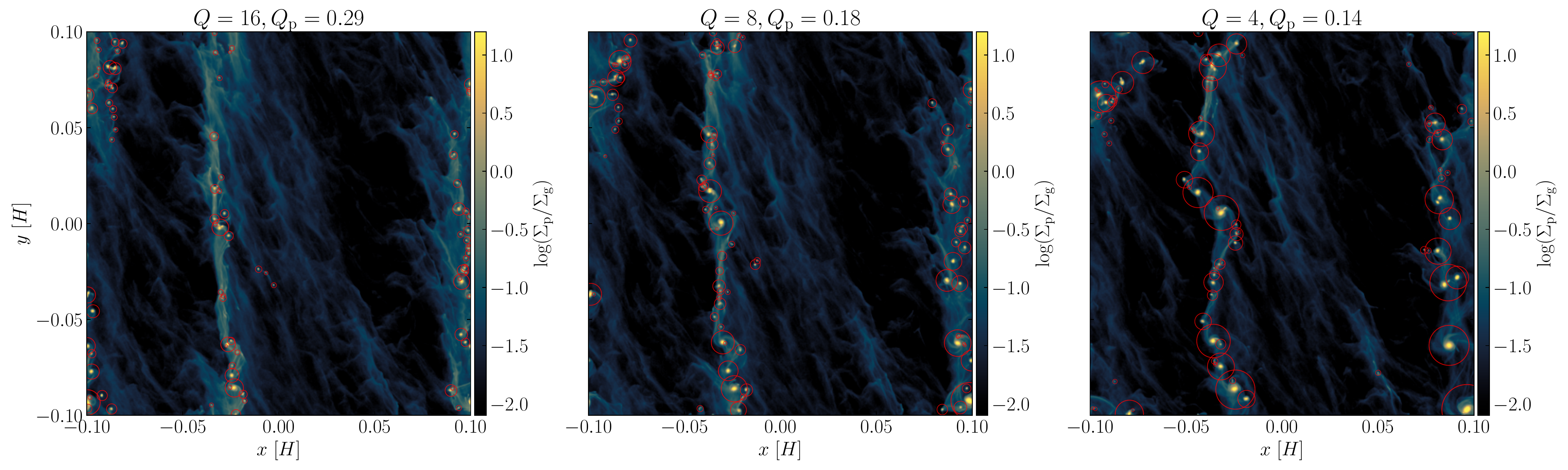}
    \caption{Snapshots of the three self-gravity simulations at $62 \Omega^{-1}$, which is $2\Omega^{-1}$ after self-gravity is turned on. All three simulations formed gravitationally bound clumps indiciated by red circles.}
    \label{fig:pls_map}
\end{figure*}

Table~\ref{tab:sim_values} summarizes the simulation parameters, the measured quantities $\delta_r$, $\delta_z$ and local particle concentration $Z$, as well as the resulting predicted properties, namely $Q_\mathrm{p}$ as well as characteristic size and mass scales. All three simulations have $Q_\mathrm{p} < 1$ and are thus expected to be gravitationally unstable. This is unsurprising as the simulation parameters were chosen in line with past setups proven to produce planetesimals \citep[e.g.,][]{Li2019}. Our smallest resolved scale is $\Delta x = 3.9 \cdot 10^{-4} H$, which is about an order of magnitude smaller than $\lambda_\mathrm{min}$ for our simulations. We thus expect to capture the small scales of diffusion regulated collapse in these simulations. On the other hand, the domain size is $0.2 H$, which is of order $\lambda_\mathrm{max}$. Moreover, the largest scale is further limited by the maximum extent of over-dense regions, which radially and vertically is of order $\eta r = \Pi H $ \citep[][]{Gerbig2020}, which for $\Pi = 0.05$ is just $5 \cdot 10^{-2} H$, and thus almost an order of magnitude less than $\lambda_\mathrm{max}$. Hence, we do not expect to the simulation to collapse on the largest unstable scales and produce planetesimals on the very high mass end. 

Fig.~\ref{fig:pls_map} shows snapshots of all three runs at $ 62 \Omega^{-1}$ which is $2.0 \Omega^{-1}$ after self-gravity is turned on. In addition to the practical benefit of reducing computational costs, we chose this  short time for self-gravity to act, in order to capture a relatively pristine mass distribution that is devoid of mergers. We used the clump finder algorithm \texttt{PLAN} \citep{Li2019Plan, Li2019} to identify gravitationally bound clumps 
, which are produced in all three runs. Indeed, all runs collapse rapidly enough for diffusion measurements to be unfeasible post 60 $\Omega^{-1}$. Thus, we cannot confirm whether or not diffusion increases in this gravo-turbulent state as it does in \citet{Klahr2021}.

Note, that the requirement of Hill-density imposes a physical mass unit onto the analytic prediciton, and only implicitly depends on distance to the star \citep[see][for a discussion on the radial dependence of the diffusion-limited collapse criterion]{Klahr2020}. On the other hand, the numerical results, i.e. the densities of bound clumps in Fig.~\ref{fig:pls_map} are fully scale-free in gas surface density $\rho_\mathrm{0}$. However, the two can be connected using the self-gravity parameter. More specifically, the mass unit of the simulation $M_0$ can be expressed as
\begin{align}
\label{eq:mass_unit}
    M_0 &= \rho_0 H^3 = h^3 \Tilde{G} \frac{M_\odot}{4\pi} = \frac{1}{4\sqrt{2\pi^3}}\frac{h^3}{Q}M_\odot \\
    &= 5.21 \cdot 10^{2} \left(\frac{Q}{16}\right)^{-1} \left(\frac{h}{0.05}\right)^{3} M_\mathrm{Ceres},
\end{align}
which introduces the same cubic dependence on aspect ratio as the analytic predictions in e.g., Eq.~\eqref{eq:massathill}. The aspect ratio relates to the pressure gradient parameter via
\begin{align}
    h = \left(-2\eta \frac{\dif \ln \rho}{\dif \ln r}\right)^{1/2} = -2 \Pi \frac{\dif \ln \rho}{\dif \ln r}.
\end{align}
Throughout this work, we choose $h = 0.05$, which together with our simulation parameter of $\Pi = 0.05$ implicitly assumes $\dif \ln \rho/ \dif \ln r = - 1/2$. These choices are relatively generic for disk models \citep[compare to e.g.,][]{Dullemond2007, Gerbig2019, Gerbig2022}.

The smallest mass above which a planetesimal is well resolved is the mass associated with a Hill-radius that equals the cell-size, i.e., $m_{r_{\rm H}=\Delta x}$. This limitation is caused by the accuracy of the self-gravity solver set by the finite grid scale \citep{Simon2016}.  \texttt{PLAN} automatically discards clumps, if any, less massive than $m_{r_{\rm H}=\Delta x}$. Using Eq.~\eqref{eq:hill_radius} and the mass scaling in Eq.~\eqref{eq:mass_unit}, this mass limit is independent of $Q$, i.e.
\begin{align}
\begin{split}
m_{r_{\rm H}=\Delta x} &= \frac{9}{4\pi}\rho_\mathrm{H}\Delta x^3  = \frac{9}{4\pi} \left(\frac{\rho_\mathrm{H}}{\rho_0}\right)\left(\frac{\Delta x}{H}\right)^3 M_0  \\
& = \frac{81}{64} h^3 \left(\frac{\Delta x}{H}\right)^3 M_\odot = 1.97 \cdot 10^{-5} M_\mathrm{Ceres}
\end{split}
\end{align}
for $\Delta x = 3.9 \cdot 10^{-4} H$.  As shown in Table \ref{tab:sim_values}, $m_{r_{\rm H}=\Delta x}$ is smaller than $m_\mathrm{p,min}$, and $\lambda_{\rm min}$ is larger than $2\Delta x$ for all $Q$, suggesting we sufficiently resolve the smallest expected planetesimals.

\subsection{IMFs and K-S test}
\label{sect:kstest}

\begin{figure*}
    \centering
    \includegraphics[width = \linewidth]{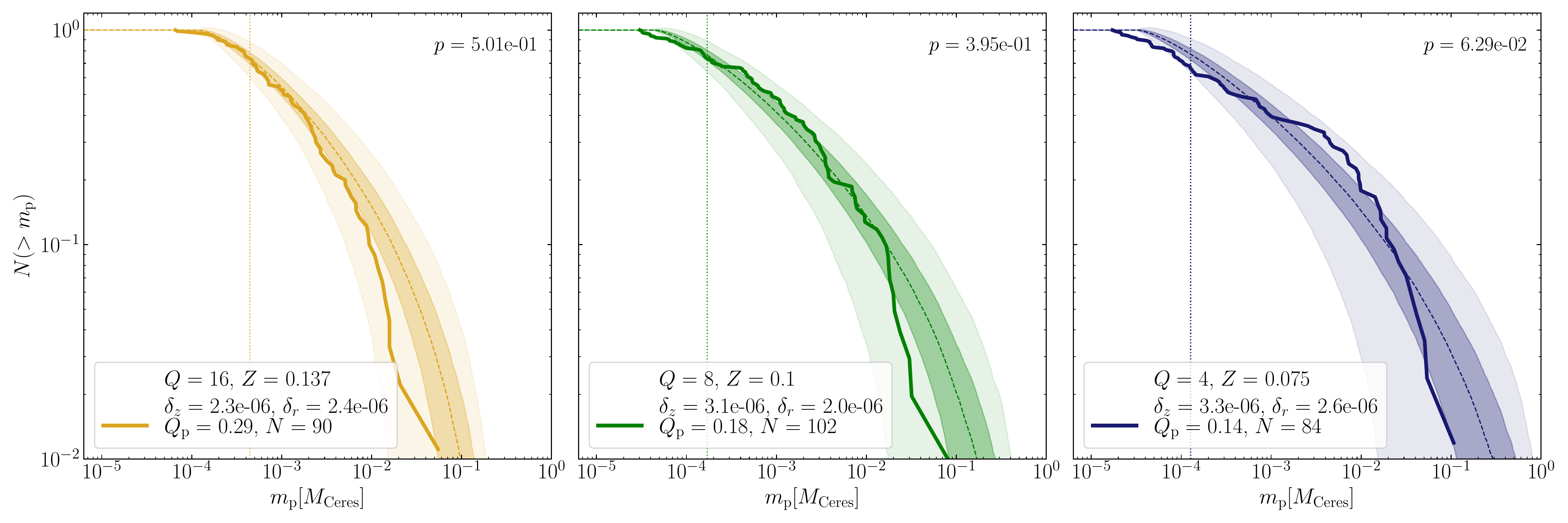}
    \caption{Initial mass functions $N(>m_\mathrm{p})$ for the three self-gravity runs. The numerically obtained IMFs are shown in solid lines and are based on the snapshots seen in Fig.~\ref{fig:pls_map}. The IMF resulting from the growth rate-based PDF are shown in dashed lines. The dark and light shaded regions indicate one and three standard deviations from the distribution of IMFs that originates from only drawing $N \sim 100$ planetesimals from the distribution. The vertical dotted lines indicate the masses associated with the fastest growing mode $m_\mathrm{p,fgm}$. Lastly, the $p$-value from the K-S-Test is shown in the top right corner of each panel.}
    \label{fig:imfs_comparison}
\end{figure*}

\begin{figure*}
    \centering
    \includegraphics[width = \linewidth]{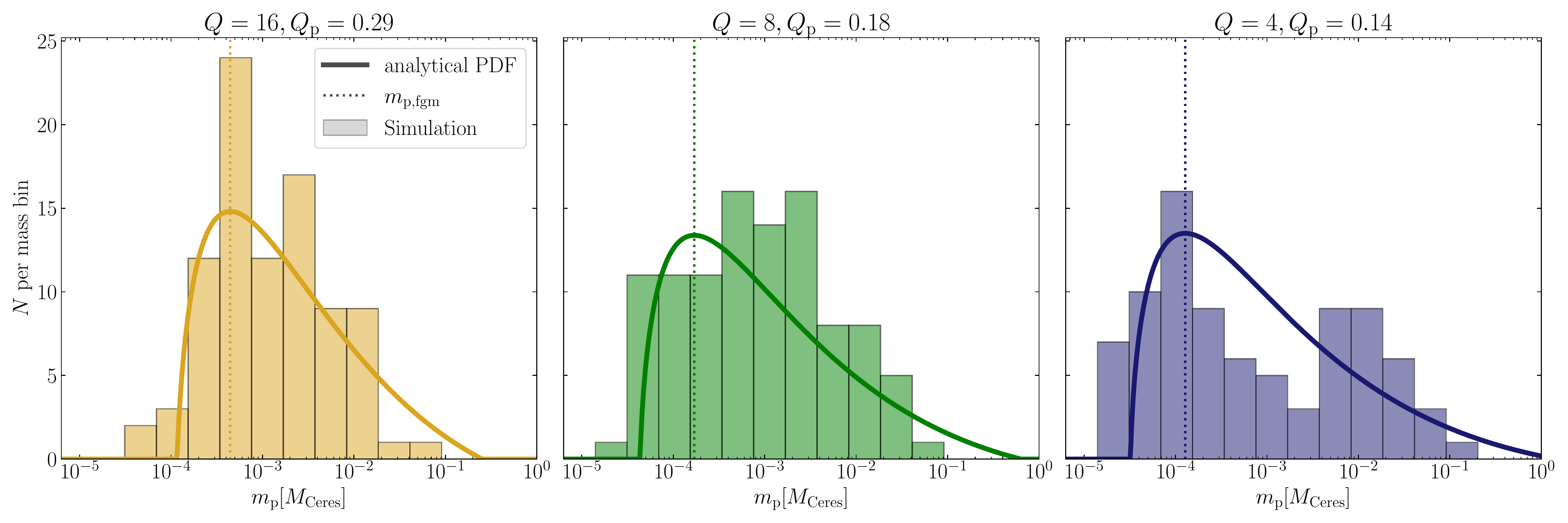}
    \caption{\new{Differential initial mass functions for the three self-gravity runs (compare to Figs.~\ref{fig:pls_map} and \ref{fig:imfs_comparison}). We show the number of clumps formed in each logarithmic mass bin as in similar figures in \citet{Li2019} in comparison to the analytical PDF we calculate (solid lines). The mass associated with the fastest growing mode (vertical dotted line) approximately corresponds to the turnover in the differential initial mass function.}}
    \label{fig:diff_imfs_comparison}
\end{figure*}

Figure~\ref{fig:imfs_comparison} shows the IMFs obtained from the snapshot \new{depicted} in Fig.~\ref{fig:pls_map} in solid lines. We compare these numerically obtained IMFs, to the predicted IMF from the Toomre-like instability paradigm, evaluated for the properties shown in Tab.~\ref{tab:sim_values}. Hereby, we draw $N$ planetesimals from the theoretical PDF, where $N$ is the number of bound clumps found in the respective simulation. We conduct this random draw 1000 times in order to obtain an average theoretical IMF (dashed curves), together with one and three $\sigma$ intervals (dark and light shaded regions respectively). In addition, we draw a dotted vertical line, corresponding to the mass associated with the fastest growing mode of the IMF prediction. 

In order to assess the quality of the analytical predictions, we perform two sample Kolmogorov-Smirnov (K-S) tests, which quantify the likelihood $p$ of two samples (in our case the numerical and analytically obtained planetesimal masses) originating from the same underlying distribution. The null hypothesis, that is the two samples indeed coming from the same distribution, can be rejected if $p < 0.05$. For the growth-rate based PDF, the K-S test yields $p$-values of $0.501, 0.395$, and $0.0629$ for $Q = 16, Q = 8$, and $Q = 4$ respectively. The numerical IMFs are therefore consistent with the analytical IMFs obtained from the Toomre-like analysis. 

\new{In Fig.~\ref{fig:diff_imfs_comparison}, we compare the differential mass distribution for our three runs to the analytical prediction for the underlying PDF. The turnover of the differential mass distributions approximately corresponds to the masses associated with the fastest growing mode, i.e. the peaks of the PDFs. For $Q=4$, the distribution is bimodal with another, smaller peak toward the higher mass end, presumably due to the stronger self-gravity where large clump mergers may already be happening when the planetesimals were counted. This deviation is reflected in the relative low $p$-value for $Q =4$ (see Fig.~\ref{fig:imfs_comparison}).}

\section{Discussion}
\label{sect:discussion}

We developed a framework of analytically obtaining planetesimal IMFs, based on the Toomre-like instability in the particle layer in protoplanetary disks. In Sect.~\ref{sect:numerical_tests} we presented a simulation of the streaming instability, measured diffusivity and particle concentration for three self-gravity runs, and then compared resulting $Q_\mathrm{p}$-values and corresponding analytical IMFs to the numerically obtained IMFs. We find that the IMFs are consistent with each other, as seen Fig.~\ref{fig:imfs_comparison}. Specifically, our prediction can explain a number of key properties of the planetesimal IMF, which we will outline in the following.

Our IMF prediction explains the somewhat counter-intuitive property of the least massive clumps being present in the simulation with the most mass, i.e. $Q =4$. Since this simulation has the smallest $Q_\mathrm{p}$ with $Q_\mathrm{p} = 0.14$, the fastest growing mode shifts to smaller scales (see Fig.~\ref{fig:disprel}) and thus the most likely planetesimal becomes less massive. 

Next, we observe that the IMF becomes flatter if $Q$ and therefore $Q_\mathrm{p}$ decreases, which is due to the increase of the range of unstable scales. Notably, all three numerical IMFs fall short of the analytical prediction on the high mass end. As alluded to in Sect.~\ref{sect:num_imfs}, this is expected as these planetesimals would have to form from scales that exceed the maximum scale of particle over-densities $\eta r$ \citep[see][]{Gerbig2020}. Indeed, the idea that the most massive planetesimal is not set by the largest Toomre-scale $\lambda$ but by $\eta r$, is consistent with our simulations where the most massive clump has $\sim 0.1 M_\mathrm{Ceres}$ in all three simulations, and also with Fig.~10 in \citet{Simon2016}, where the most massive planetesimal shortly after self-gravity is turned on is relatively independent of $\Tilde{G}$ (or equivalently $Q$). 

Lastly, the fact that $\lambda_{\rm fgm}$ is much closer to $\lambda_{\rm min}$ when $Q_{\rm p}$ is small, provides an explanation for why only high-resolution simulations can sufficiently capture the turnover in the mass frequency distribution at the low-mass end of planetesimal IMFs \citep{Li2019}. 

\subsection{Caveats}
\label{sect:caveats}

In this section, we outline a number of caveats that are to be kept in mind when relating the herein presented theory and analysis to real disks. 

First, unstable scales $\lambda$ were converted to planetesimal masses by assuming the contraction of a circular \new{region} of radius $\lambda/(2\pi)$ at Hill-density. While such a conversion agrees with timescale arguments presented in e.g. \citet[][]{Gerbig2020, Klahr2020}, it certainly is an order of magnitude calculation. In particular, as pointed out in \citet{Polak2022}, a sphere at Hill-density is not fully tidally stable. The Hill-sphere is derived by considering the tidal forces on either end of the Roche lobe in the restricted three-body potential. This has the Hill-sphere extend beyond the tidally-stable Roche lobe, and implies Hill-density is insufficient for tidal stability by an order of unity factor. Moreover, it is not the case that all unstable scales are at Hill-density exactly. In fact, as evidenced by Fig.~\ref{fig:diffusion_measurebig}, many particles are located within regions which exceed Hill-density by up to an order of magnitude.

Additionally, our model assumed planetesimal formation to be perfectly efficient. Fig.~\ref{fig:pls_map} however demonstrates that there are still particles in the filaments that are not bound to any clump.  Moreover, previous works showed that higher resolution simulations may extend the mass distribution to lower masses \citep[][]{Li2019}. Another possibility that is not considered in our model is a single clump forming binary planetesimals upon contraction to material density \citep[][]{Nesvorny2019, Nesvorny2021}. 

We note, that while multiplicity and inefficient contraction would shift the IMF towards smaller masses, collapse beyond Hill-density and double counting would shift the IMF towards larger masses. We expect that this allows the numerically obtained IMFs to remain consistent with our analytical predictions.

Other simplifying assumptions include gas density being considered constant. For stable, yet small $Q$ values, the gas develops a mid-plane cusp with a non-neglible vertical density gradient \citep[see e.g.,][for a review]{Armitage2015}. \new{Our simulations are conducted in quiescent disks without an external source of turbulence in the gas phase. Gas turbulence can increase particle diffusivity which is manifested in turbulent thickening of the particle mid-plane \citep{Gole2020}. This increases planetesimal masses (see Eq.~\eqref{eq:massathill}), but at the same time, imposes higher demands on the particle concentration $Z$ for achieving an unstable $Q_\mathrm{p} < 1$ in the first place.} We also assumed a mono-disperse dust population with a constant Stokes number. This choice is likely reasonable to first order as in real disks, most solid mass is contained in the largest grains   \citep[e.g.,][]{Birnstiel2011, McNally2021}, and the largest grains dominate dust-gas interactions in the mid-plane and participate most vigorously in planetesimal formation \citep{Yang2021}. Still, a multi-species dust fluid can alter the turbulent behavior of the system itself \citep[see e.g.,][who pointed out the damping effect multi-species dust can have on the streaming instability]{Krapp2019, Schaffer2021}. Therefore, the assumption of single-species dust is to be kept in mind when applying our results to real disks. 

Lastly, we acknowledge that our numerical setup does not perfectly replicate the assumptions the Toomre-like instability is based on. More specifically, the Toomre analysis, assumes a thin, two-dimensional disk of constant density which is then linearly perturbed in form of axisymmetric waves. This picture evidently differs from the turbulent state, i.e. non-linear streaming instability in vertically-stratified disks, our simulations are in when self-gravity is turned on. We attempted to mitigate this discrepancy by evaluating diffusivities and concentration locally, rather than globally, which, as discussed in Sect.~\ref{sect:kstest}, yielded consistent IMFs. 

\subsection{The shape of the initial mass function in real disks}

The diffusion-tidal-shear collapse paradigm suggests that the shape of planetesimal IMF depends on the gravitational, Toomre-like instability and on the stability parameter $Q_\mathrm{p}$ only. While independent information on diffusivities $\delta_r$ and $\delta_z$, as well as Stokes number $\mathrm{St}$, and disk aspect ratio $h$ are required to match the IMF to the appropriate mass range, they do not affect the steepness of the IMF (for a fixed $Q_\mathrm{p}$). As such, as a disk region collects more massive particles, it experiences an increase in average grain size, cools or becomes less turbulent, its $Q_\mathrm{p}$-value decreases, thus flattening the planetesimal IMF. Within this context, the role of the streaming instability (or other processes affecting particle dynamics) for the planetesimal IMF in this work, is to set the initial condition at the onset of gravitational collapse. The key question within our framework therefore is to determine what value for $Q_\mathrm{p}$ takes on in real disks. 

Our simulations, much like many in previous works \citep[e.g.,][]{Simon2016, Schaefer2017,Li2019, Gerbig2020} show collapse as soon as self-gravity is turned on. The turbulent state before self-gravity is turned on is therefore not physical, and thus the $Q_\mathrm{p}$-values and IMFs we calculate not necessarily reflective of real disks. Indeed, as discussed in \citet{Gerbig2020}, if a system softly evolves towards instability --- for example by collecting more particle mass --- its $Q_\mathrm{p}$-value likewise softly evolve from $Q_\mathrm{p} > 1$ to $Q_\mathrm{p} < 1$. If $Q_\mathrm{p} \sim 1$ is sufficient for the system to fully collapse and form planetesimals as assumed in \citet{Klahr2020, Klahr2021}, then the resulting IMF is expected to be rather steep \citep{Polak2022}. On the other hand, the system may be unable to fully contract at $Q_\mathrm{p} \sim 1$ if it is below Hill-density. To a similar effect, as the contraction time, given by \citep{Gerbig2020},
\begin{align}
    t_\mathrm{c} = \frac{\Omega}{4\pi G \rho_\mathrm{p}\mathrm{St}} = \sqrt{\frac{\pi}{8}}\sqrt{\frac{\delta_z}{\delta_r}}\frac{Q_\mathrm{p}}{\mathrm{St}} \Omega^{-1},
\end{align}
only falls below the dynamical timescale $\Omega^{-1}$ once $Q_\mathrm{p}$ is substantially less than unity, it is plausible that the system can collect relevant amounts of mass during contraction. Both situations would lead to smaller values for $Q_\mathrm{p}$ and thus flatter IMFs like the ones in our simulations in Fig.~\ref{fig:imfs_comparison}.

One observational constraint may be provided by the Asteroids size distribution, which, at least in part, constitutes pristine planetesimals from the Solar System's formation era \citep[see][for a review]{KlahrDelboGerbig2022}. Taking, the collisional evolution into account, it is possible to reconstruct model of the primordial size distribution of Asteroids \citep[see e.g.,][]{Delbo2019}. Within the here presented framework, the ``knee'' at $\sim 100$ km \citep[e.g., ][]{Gladman2001} typically found in the resulting mass functions can be interpreted as an indication of a marginally unstable origin system with $Q_\mathrm{p}$ close to unity --- that is assuming the asteroids in question indeed formed via gravitational collapse of locally over-dense regions, which very well may not have been the case. Cold Classical Kuiper Belt Objects (KBOs) are believed to be even more primordial with little to no collisional evolution  \citep{Morbidelli2020}. \citet{Kavelaars2021} found \new{an} exponentially tapered power law as the best fit for the Cold Classical KBO mass distribution, and connected the lack of large planetesimals ($>$ 400 km in size) to streaming instability regulated planetesimal formation. Our work is consistent with this interpretation, as $Q_\mathrm{p}$ imposes a limit on the maximum planetesimal mass.

One often invoked pathway of forming planetesimals and circumventing the meter barrier, that is worth discussing in the context of our work, is the existence of long-lived pressure bumps associated with disk sub-structures. While such structures are robustly confirmed observationally \citep[e.g.,][]{Andrews2018}, their role in producing the first generation of planetesimals is elusive, as already existing planets seem to most promisingly explain the sub-structures in the first place \citep[][]{Teague2021}. Either way, a pressure bump is characterized by very small pressure gradients, which diminishes the relative velocity between dust and gas. If $\Pi = 0$, then particles can trivially settle razor-thin and go gravitationally unstable \citep[as seen in e.g.,][]{Abod2019}. In this state, both vertical and radial diffusivities are expected to trend towards zero, implying $Q_\mathrm{p} \rightarrow 0$ as well, which our work associates with the planetesimal IMF shifting towards very small masses.  Unless hierarchical mergers between small planetesimal immediately after formation is very efficient, such a bottom-heavy mass function seems unlikely. We therefore suspect that either the majority of planetesimal formation does not occur in $\Pi = 0$ regions; or, that other mechanisms, possibly gravito-turbulence \citep[see e.g.,][for the gas disk]{Riols2017}, provide a lower bound on diffusivities, thus also setting a lower limit on $Q_\mathrm{p}$.

To conclude, our analysis that connects the PDF of formed planetesimals to the growth rates of the Toomre-like instability of a particle layer subject to a diffusive flux, unifies both the flat, power-law shaped, IMFs previously obtained numerically \citep[e.g.,][]{Simon2016, Li2019}, and the `Asteroids are born big' \citep{Morbidelli2009} paradigm that arises when investigating marginally unstable systems \citep{Klahr2020}. The analytically obtained IMFs are consistent with our numerical setups. Further work is required to test the predictions for different numerically setups, and in particular, to assess how the value of $Q_\mathrm{p}$ at the on-set of planetesimal formation, and thus the steepness of the resulting IMF, depend on disk properties and radii. Such an analysis informs initial conditions for planet formation models \citep[e.g.,][]{Emsenhuber2021}, and conversely, if the formation of an observed exoplanetary population presupposes a specific planetesimal IMF\citep[e.g.,][for rocky Super-Earths]{Batygin2023}, provides constraints on the disk state during the era of planetesimal formation.

\section{Acknowledgements}
\label{section:acknowledgements}

\new{We thank the referee for insightful comments that improved the quality of the manuscript.} This work and KG and RL benefited from the 2022 Exoplanet Summer Program in the
Other Worlds Laboratory (OWL) at the University of California, Santa
Cruz, a program funded by the Heising-Simons Foundation. The authors thank Greg Laughlin,  Hubert Klahr, Andrew Youdin, Ruth Murray-Clay and Malena Rice for insightful comments and discussions.

\software{Athena \citep[][]{Stone2008}, PLAN \citep{Li2019Plan}, NumPy \citep{Harris2020}, Matplotlib \citep{Hunter2007}, CMasher \citep{vanderVelden2020}}

\bibliography{references}{}
\bibliographystyle{aasjournal}

\end{document}